\documentclass{aastex}
\pdfoutput=1

\shortauthors{Ribeiro et al.}

\begin{document}


\title{Activity on the M star of QS Vir.}


\author{Ribeiro, T.}
\affil{Departamento de F\'isica, Universidade Federal de Santa Catarina, Campus Trindade, 88040-900 Florian\'opolis, SC, Brazil}

\author{Kafka, S.\altaffilmark{1}}
\affil{Department of Terrestrial Magnetism,Carnegie Institution of Washington, 5241 Broad Branch Road NW, Washington, DC 20015}

\and

\author{Baptista, R}
\affil{Departamento de F\'isica, Universidade Federal de Santa Catarina, Campus Trindade, 88040-900 Florian\'opolis, SC, Brazil}

\and

\author{Tappert, C.}
\affil{Departamento de Astronom\'ia y Astrof\'isica, Pontificia Universidad Cat\'olica, Vicu\~na Mackenna 4860, 782-0436 Macul, Chile}


\altaffiltext{1}{email: tiago@astro.ufsc.br}


\begin{abstract}

We report analysis of VRIJH photometry, and phase-resolved optical spectroscopy  of the eclipsing DA white dwarf plus dMe dwarf binary QS Vir. Modeling of the photometric data yields an inclination of $i = 74.9\pm0.6$ and a mass ratio of $q = M_2/M_1 = 0.50\pm0.05$. Our Doppler maps indicate the presence of material in the Roche lobe of the white dwarf, at a location near the M star, likely due to accretion from the stellar wind of the M star (as opposed to Roche-lobe overflow accretion).  We also constructed images of the brightness distribution of the M star at different epochs which reveal the location of two stable active regions. Doppler tomography shows that the majority of the Hydrogen and Ca~II H\&K emission originates on the active M dwarf, likely distributed in two preferred activity longitudes, similar to active regions on BY Dra and FK Comae systems.

\end{abstract}


\section{Introduction}
\label{sec:intr}

The progenitors of close binary stars with a white dwarf as primary (i.e., more massive) component are thought to be initially wide detached binaries. In the course of its nuclear evolution, the more massive star expands first. If the separation is sufficiently small, this will yield a Common Envelope configuration, where both the remaining core of the more massive star and its lower mass companion revolve around each other within the expelled material from the primary. Responding to the enhanced angular momentum loss due to friction, the separation between the two stars decreases, and when the envelope material is expelled as a planetary nebula, the aforementioned close binary emerges. Such systems are known as post common envelope binaries (PCEB). 

Angular momentum is continuously lost via gravitational radiation and via a magnetically confined stellar wind from the main sequence star component (\textit{magnetic braking}). As a result, the orbital period (and the binary separation) decreases with time, until the lower mass dwarf eventually fills its Roche lobe. Mass transfer then starts through the inner Lagrangian point (L1) of the binary, dumping material onto the white dwarf (WD) via an accretion disk or the magnetic field lines of the  white dwarf.  When the dwarf is a K/M star, those systems are known as cataclysmic variables (CV)\footnote{For a comprehensive description of CVs the reader is referred to \cite{c01} and \cite{war}.}. The orbital period (and the binary separation) at which first contact occurs depends strongly on the radius/mass of the low mass main sequence star (secondary) and on the mass of the WD (primary). If the time required for the secondary star of a PCEB to come into contact with its Roche lobe is shorter than the Hubble time ($\sim 10$Gyr) the system is called a pre-CV \citep{S_G_2003}.

\defcitealias{2003MNRAS.345..506}{DOD03}

QS Vir (or EC 13471-1258) is an eclipsing WD+M dwarf binary system, discovered in the Edinburgh-Cape Blue Object Survey \citep{1997MNRAS.287..848S}. \citet{2002AJ....124.2853K} and  \citet[hereafter \citetalias{2003MNRAS.345..506}] {2003MNRAS.345..506} provided an extensive study of the system, combining photometric monitoring, time-resolved spectroscopy, and (in the latter case) UV spectra. By measuring the rotational velocity of the WD \citetalias{2003MNRAS.345..506} shown that it is consistent with that of accreting CV primaries \citep{szk}. Therefore, they proposed that the system is not a pre-CV, but instead a hibernating CV; i.e. a former CV that became detached as the consequence of a nova eruption \citep{1986ApJ...311..172P,1986ApJ...311..163S}. Their photometric monitoring for over 10 years revealed small orbital period deviations from a linear ephemeris which they suggest is a result of random variation in the orbital period of the system, possibly caused by activity on the M secondary. A number of optical flares were observed, confirming that the M star is magnetically active. Phase-resolved optical spectra suggested that low level accretion is actually taking place in the system.

We present here additional time-resolved blue spectra of the system using the RC Spectrograph at the CTIO $4.0$m Blanco Telescope and Echelle spectra taken with the VLT, as well as recent near-infrared photometric data acquired with the Brazilian $1.6$m Perking \& Elmer telescope at Pico dos Dias.  We also analyze the R and V(RI)c light curves of  \citet{2002AJ....124.2853K} and \citetalias{2003MNRAS.345..506}.  We use a light curve inversion method and Doppler tomography to explore the active regions on the M star, providing a pictorial guide to their location and evolution. This paper is organized as follows. In section \ref{sec:data_red} we present data acquisition and reduction procedures, while in section \ref{sec:anl} we report the results from the analysis of the data with the different tools. We follow with a discussion of the combined analysis and a comparison with previous results in section \ref{sec:disc}. Final remarks conclude our paper in section \ref{sec:end}.

\section{Data acquisition and reduction}
\label{sec:data_red}

\subsection{Near-IR Photometry}
\label{subsec:obs_nir}
NIR photometry of QS Vir was collected with the CAMIV infrared camera attached to the 1.6 m Perkin \& Elmer telescope of the Observat\'orio do Pico dos Dias (OPD/LNA), in southern Brazil, on 2008 March 18. CAMIV has a $1024 \times 1024$ pixels HAWAII detector cooled to 77K,  designed for observations at $0.8-2.4 \micron$ (for further details see \citealt{2006MNRAS.369..1972}). The sky was patchy at the night of the observation and the humidity was very high, interrupting the observations before completion of a full orbit. Fast wind was also blowing at the dome, tilting the telescope in some cases. A log summarizing the observations is presented in Table \ref{tab:obs_log}.

In order to avoid cycle to cycle brightness variations (very common in QS Vir, \citetalias{2003MNRAS.345..506}) the photometry was performed in a quasi-simultaneous mode, alternating between J and H filters after each dithering sequence. Data reduction was performed with CIRRED/IRAF\footnote{IRAF is distributed by the National Optical Astronomy Observatories, which are operated by the Association of Universities for Research in Astronomy, Inc., under cooperative agreement with the National Science Foundation.}. The non-linearity, inherent to all reversed polarized semiconductor detectors, was corrected using the task IRLINCOR with the linear coefficient provided by \cite{2004PhD_INPE}. A mask for bad pixels was constructed using a normalized median of flat-field images. Sky level and dark frame subtraction as well as flat-field division and bad pixel mask corrections were applied to each frame.

Aperture photometry was then extracted for the target star and $5$ other nearby field stars, with the brighter field star being used as a reference star for differential photometry. This procedure minimizes the effects of sky transparency variations along the night and is particularly useful for non-photometric nights -- such as that of our observations. In order to keep track of the changing position of the stars in the field due to the dithering procedure, we have used the task IMMATCH/CTIO. The relative magnitude of the target was converted to absolute flux (in $mJy$) using the absolute magnitude of the reference star together with zero point constants provided by the \textbf{2MASS} project \citep{2006AJ.131..1163}. The data was phase-folded according to the mid-eclipse ephemeris of \citetalias{2003MNRAS.345..506},
\begin{equation}
\label{eq:efem}
{\rm T_{mid}} ({\rm E}) = {\rm HJD } ~2~448~689.64062 + 0.150757525 \cdot {\rm E}
\end{equation}
where E is the eclipse cycle.

\subsection{Spectroscopic Observations}
\label{subsec:obs_spec}
Most of our spectroscopic observations were obtained with the RC Spectrograph on the CTIO 4-m Blanco telescope on 2007 March 20. We obtained 46 spectra using the KPGL3 grating, yielding a spectral range from 3800-7500\AA~at 1.2\AA/pixel, which provided a resolution of $\sim$3\AA. A log of our observations is included in Table \ref{tab:obs_log}. For detector calibrations, we used the standard IRAF procedures, and for data reductions we used IRAF's ONEDSPEC/TWODSPEC packages. Due to non-photometric conditions (some passing clouds) we did not obtain spectroscopic standards, therefore our data are not flux calibrated. Example spectra are presented in Fig. \ref{fig:main_spec}.

The Echelle spectrum was taken as a 30 min exposure on 2007 August 16 with UVES at UT1, ESO-Paranal. The DIC2 mode was used with cross-dispersers \#2 and \#4 and central wavelengths 3900 {\AA} and 7600 {\AA}. Reduction of the data was performed using the Gasgano interface with the ESO/UVES pipeline scripts. This included Bias subtraction, flat-fielding, wavelength calibration, and the correction for the instrumental response function and atmospheric extinction using flux standard LTT 7987. The output consists of three spectra with useful data in wavelength ranges 3800--4510 {\AA}, 5720--7480 {\AA}, and 7700--9430 {\AA}, respectively. The spectra have slightly different resolving powers around $R \sim 70\,000$. 

\section{Data analysis and discussion}
\label{sec:anl}

\subsection{Photometry}
\label{subsec:anl_nir}

\begin{figure}
\epsscale{1.0}
\plotone{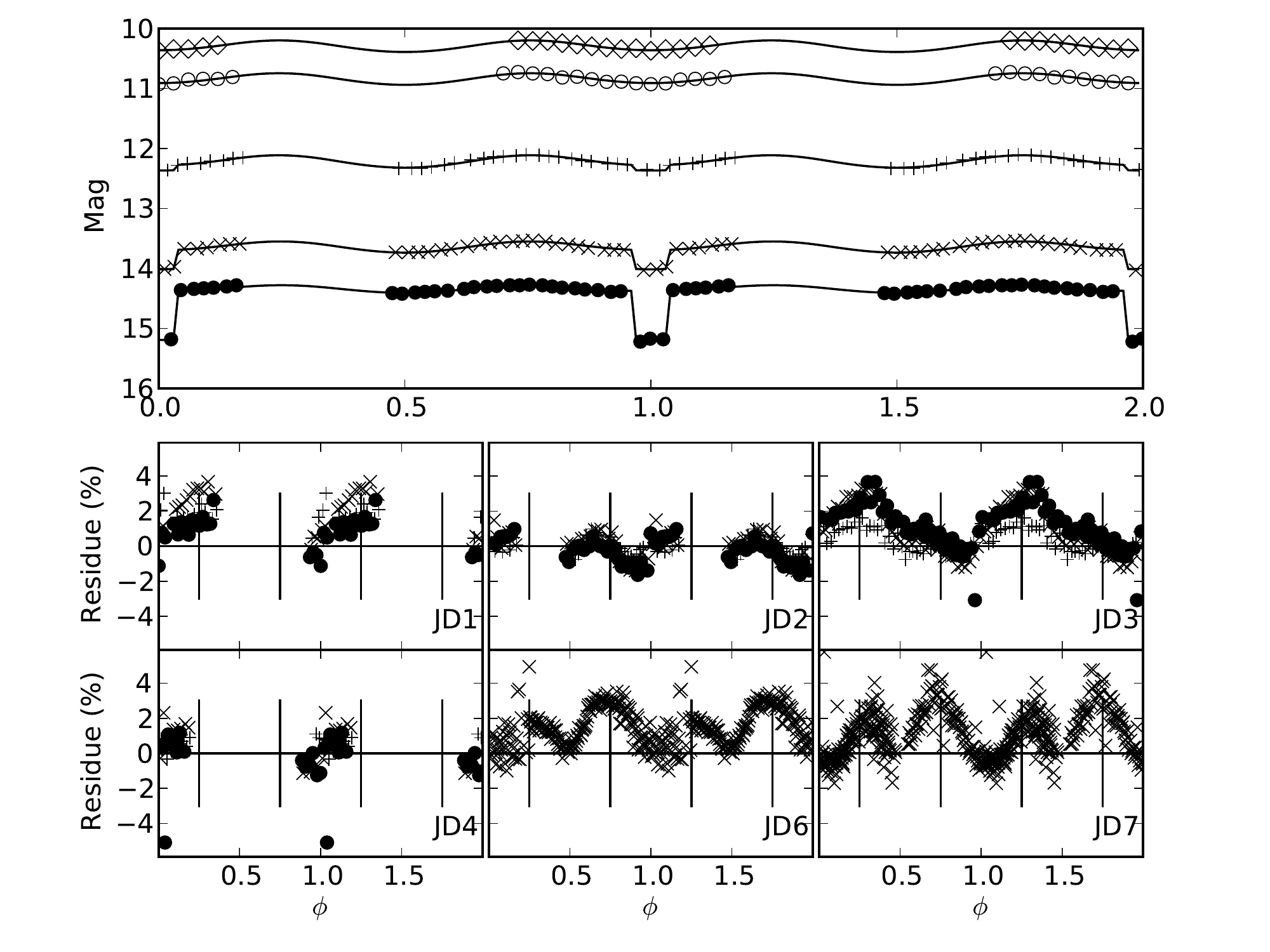}
\caption{The upper panel shows our NIR light curves and JD2 data from \cite{2003MNRAS.345..506} of QS Vir. Model light curves are shown as solid lines and different symbols are used for each filter. From top to botton, the  NIR light curves are shown as open diamonds (H) and open circles (J) and the JD2 are in plus signs (I), crosses (R) and filled circles (V). The phases are repeated for better visualization. In the six bottom panels the residue shows deviations from the model (obtained for the NIR and JD2 light curves) to the VRI data of JD1, JD2, JD3 and JD4 and the R data of JD5 and JD6. The symbols are the same as in the top panel, for each filter. An horizontal line depicts level zero and vertical lines depicts phases of quadrature ($\phi$ = 0.25, 0.75,1,25 and 1.75).}
\label{fig:cl_res}
\label{fig:cl_01}
\end{figure}

The V(RI)$_c$ data of \citetalias{2003MNRAS.345..506} (see their Table 6) consist of four consecutive nights of observations in 1993. We will refer to the different sets of light curves from HJD 2\,449\,161 to 2\,449\,164 as JD1 to JD4, respectively\footnote{JD1=HJD 2\,449\,161; JD2=HJD 2\,449\,162; JD3=HJD 2\,449\,163 and JD4=HJD 2\,449\,164} and to our NIR light curves of HJD 2\,454\,544 (from 2008) as JD7. In addition, the R data from  \citet{2002AJ....124.2853K} consist of 2 cycles separated by approximately 3 days and will be referred to as JD5 and JD6. In Fig. \ref{fig:cl_01} we present our JH light curves and the JD2 V(RI)$_c$ data set from \citetalias{2003MNRAS.345..506} together with a model fit to the data. 

Fig. \ref{fig:cl_01} shows that the eclipse of the WD (primary) becomes shallower and the ellipsoidal modulations of the secondary more pronounced with increasing wavelength, i.e., the contribution of the cool secondary becomes more important than that of the compact hotter primary at longer wavelengths. In order to probe the photometry for information on the system components we have applied a light curve synthesis code similar to the \cite{wd_code} code. For a detailed description of our procedure
see \cite{2007A&A...474..213R}. In short, we assume that the surface of the secondary star is given by one gravitational equipotential within its Roche lobe and we modify its radiation field to account for gravity- and limb-darkening effects. Because the primary is assumed to be a point-like source its ingress/egress features (partial eclipse phases) are not modeled.

As a starting point to the fitting procedure, we estimated the WD flux $F_1$ at each wavelength from the depth of the eclipses and assumed that $F_1$ is constant throughout the orbit. We also obtained an initial estimate of the M star flux $F_2$ at each wavelength from the average flux during eclipse. The best-fit solution is obtained by an algorithm that searches for the set of parameters which minimizes the normalized $\chi^2$ constraint function,
\begin{equation}
\label{eq:chi2}
\chi^2 = \frac{1}{N} \times \sum_{n=1 ... N} [ \frac{F_o(\phi_n) - F_m(\phi_n)}{\sigma_o(\phi_n)} ]^2,
\end{equation}
where $F_o(\phi)$ and $F_m(\phi)$ are the observed and modeled fluxes at phase $\phi$, respectively, $N$ is the number of points in the light curve and $\sigma_o(\phi)$ is the flux error at phase $\phi$. 

The light curves are fitted simultaneously for a single set of $q$, $i$ and $r_2$\footnote{The radius of the secondary star in units of the orbital separation ($a$).} values. Since the relative contribution of each star changes with wavelength, we adjusted $F_1$ and $F_2$, the contributions of the WD and secondary respectively, accordingly in each data set. The resulting parameters (and corresponding $\chi^2$ values) are listed in Table \ref{tab:params} and were used to produce the model light curves of Fig. \ref{fig:cl_01}. The values listed in Table \ref{tab:params} are the median values obtained through $10^3$ Monte Carlo simulations, while the errors are the corresponding standard deviations. For each Monte Carlo simulation, the value of each data point in the light curves is changed according to a gaussian distribution with mean equals to the true data point and width equals to its photometric uncertainty. Each Monte Carlo light curve is then subjected to the same real data fitting procedure. Although a fraction of simulations ($\sim 20\%$) results in secondaries filling the Roche lobe, all resulting values are used to estimate the uncertainties.  Table \ref{tab:params} list the radius of the secondary ($r_2$) and the volumetric radius of its Roche lobe ($r_{\rm RL}$, as proposed by \citealt{1983ApJ...268..368E}), both in units of the orbital separation ($a$). We also show the resulting radius of the secondary in units of its Roche lobe radius ($r_2/r_{\rm RL}$). 

The values obtained by the light curve modeling procedure are in excellent agreement with those of \citetalias{2003MNRAS.345..506}, which were independently obtained by combining the measurements of eclipse width and partial phases duration with K$_1$ and K$_2$ velocities measurements. However, the resulting $\chi^2(\gtrsim 2)$ for the fit to the V(RI)$_c$ data is somewhat higher than expected for a good fit ($\sim 1.0$), showing similar deviations in the (O-C) diagram for all filters (see bottom panels of Fig. \ref{fig:cl_res}). The amplitude of the deviations in the JD2 light curves is similar in V and R, and lower by a factor of two in the I band. 
A similar analysis is performed for the JD1, JD3, JD4, JD5 and JD6 data; the results are shown at the bottom panel of Fig. \ref{fig:cl_res}. Since the JD5 and JD6 light curves are given in relative magnitudes \citep{2002AJ....124.2853K}, it is not possible to flux calibrate this data set. Therefore we use differential magnitudes for all our data in order to provide a direct comparison of the amplitude of the variations. Although different, the deviations from the model are present in all data sets. They are apparently larger for the JD1 and JD3 data and clearly larger for JD5 and JD6.

We checked whether the shape of the JD2 residual curve is a consequence of a wrong choice of system parameters. Given the symmetric shape of the ellipsoidal modulations, wrong values of orbital parameters ($i$, $q$ and $r_2$) or stellar component flux ($F_1$ and $F_2$) would produce systematic residual peaks at quadrature (orbital parameters) or conjunction (component flux). The observed residuals are different from both predictions. Alternatively, a small error in binary phase (say, $\delta \phi \sim 0.02$) could qualitatively explain the observed residuals. However, the amount of phase shift required to explain the residuals would result in a perceptible displacement of the sharp primary eclipses, which is also not observed. Therefore, the inferred systematic, phase-dependent residuals in the light curves seem to reflect intrinsic deviations from uniformity in the surface brightness distribution of the secondary star.  We will return to this point later in the paper.  
We adopted the system parameters derived from the analysis of the JD2 plus JD7 light curves (which are flux calibrated) to compute the residuals of the JD5 and JD6 data.

In Table \ref{tab:params} we list the flux for each component of the system in each filter.  We used FITSPEC/SYNPHOT to search for the best-fit blackbody spectrum to the observed WD and LMD V(RI)c JH fluxes. We infer temperatures of $T_1 = 18000 \pm 5000K$ and $T_2 = 2800 \pm 200 K$ for the primary and secondary, respectively. Furthermore, assuming radii of $R_1 = 0.011 R_\odot$ and $R_2 = 0.42 R_\odot$\citepalias{2003MNRAS.345..506} we find distances of $d_1 = 60 \pm 20$ pc and $d_2 = 40\pm5$ pc. The large uncertainty obtained for the temperature and distance to the primary star is a consequence of using essentially red wavelength coverage to constrain the temperature of blue (hot) spectra.

\citetalias{2003MNRAS.345..506} discuss the presence of a flare in their JD3 light curve at the beginning of their observations.  Similar deviations from model light curves for a non-spotted secondary star were also reported by \citet{2005Ap&SS.296..481K} (see their Fig. 1) with regard to the JD5 and JD6 data.  At the bottom panels of Fig. \ref{fig:cl_res} we show an (O-C) of the JD3 data where it is possible to check the resulting increase in flux around phase 0.25, when the system is close to quadrature.  Variations of comparable amplitude are also seen in JD5 and JD6.  Motivated by the amplitude and shape of these variations, we applied a light curve inversion technique \citep{1998AA.332..541} to assess the location of the active region(s) responsible for the flaring (in the case of JD3) and the origin of surface brightness variations that could be responsible for the deviations of the JD2, JD5 and JD6 light curves from the model. Data from JD1 and JD4 were not included in this analysis due to their incomplete orbital coverage.

We have implemented the entropy inversion procedure described by \cite{1998AA.332..541}, which has been successfully applied to data of active single M dwarf  stars, in order to derive the position and number of spotted regions \citep{2007AN....328..897K}. Since the original description of the method concerns only single spherical stars, we included the tidal distortion of the surface of the M dwarf star of QS Vir. This allows our code to reproduce the model light curves of Fig.\ref{fig:cl_01} for a secondary star with uniform brightness. We then modified the intensities of a grid of surface elements, searching for the brightness distribution that yields the best fit to the data. The procedure attempts to simultaneously minimize the reduced  $\chi^2$ of equation \ref{eq:chi2} and to maximize the entropy of the intensity grid (see \citealt{1987Obs...107...86S}) in order to select the most featureless brightness distribution that fits the data. We must also keep in mind that the one-dimensional light curves do not provide information on the latitudinal position of spotted region and that, in this case, the entropy regularization algorithm tends to place structures at the center of the projected stellar disk.

\begin{figure}

\epsscale{1.0}
\plotone{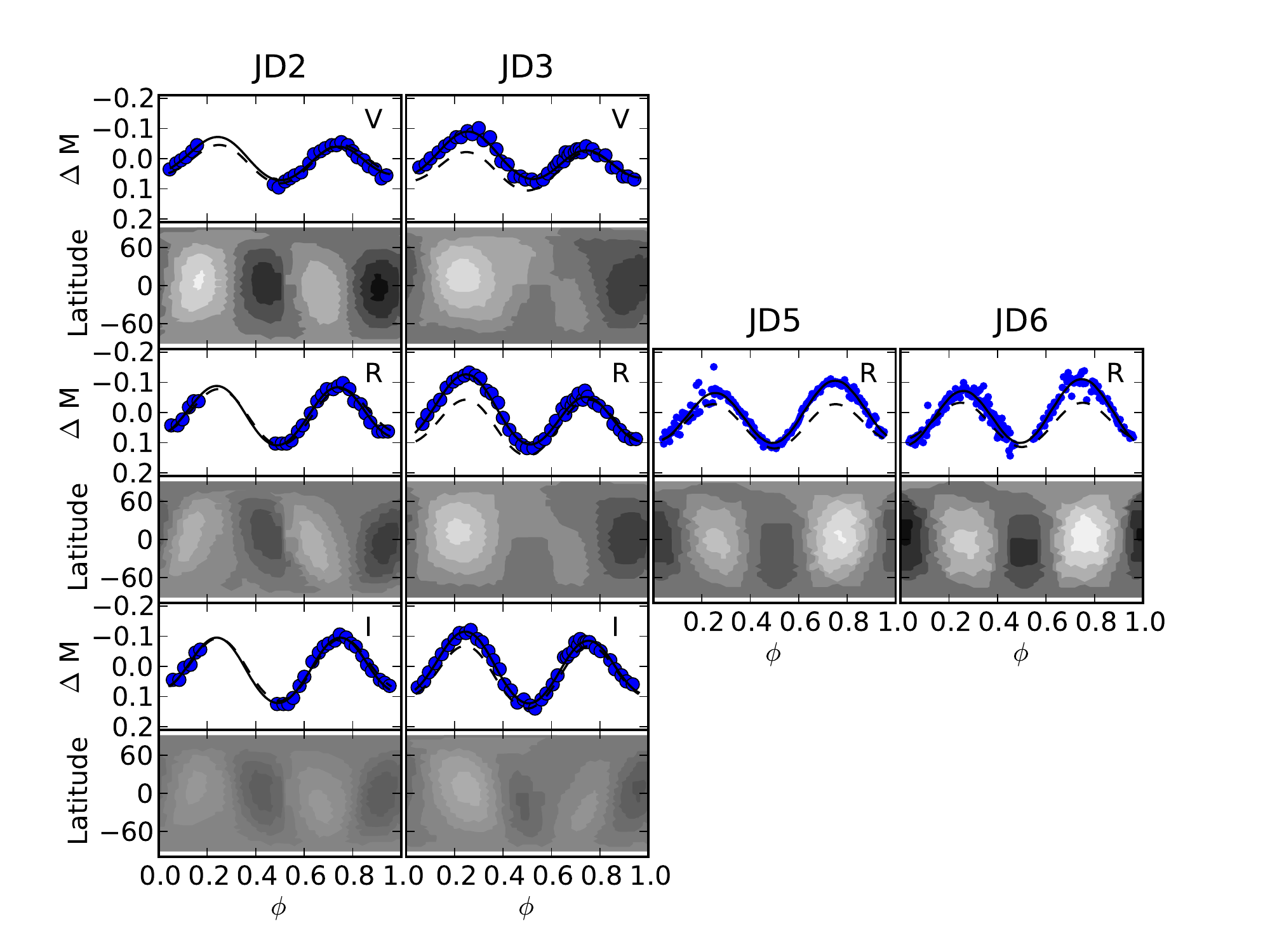}

\caption{ Application of light curve inversion method on the differential light curves of JD2 (left), JD3 (middle), JD5 (upper right) and JD6 (lower right) data. The resulting light curves are plotted as solid lines, while the light curves for a uniform brightness distribution are shown as dashed lines. Below each light curve, we present the result of the light curve inversion method. Lighter regions are brighter.}
\label{fig:clt}

\end{figure}

We show the resulting model light curves of the inversion process as solid lines in Fig. \ref{fig:clt}. Dashed lines shows the orbital modulation of an uniformly bright secondary star as a comparison. The gray scaled images presented in Fig. \ref{fig:clt} are the changes in the surface brightness distribution of the secondary required to produce the observed deviations from the orbital modulation of a distorted star with uniform brightness. All maps are shown in the same gray scale to allow for a direct comparison of different filters at different epochs. Since broad band photometry measures essentially the photosphere of the star, positive deviations from the uniform brightness model light curve (dotted lines) represent active, hot regions and will appear bright in the images. On the other hand, negative deviations represent cooler regions and will appear dark in the images.

The JD2 maps show that there are two different types of structures.  Close to phases $\phi \sim 0.2$ and $\phi \sim 0.7$ we see two regions of increased emission that are $\sim 0.5$ in phase apart. It is also possible to see two regions of decreased emission at phases $\phi \sim 0.4$ and $\phi \sim 0.9$, also $\sim 0.5$ in phase apart. These structures, which represent spotted (cool) regions, are present in all three light curves and are more evident at shorter wavelengths (V). An interesting aspect of the morphology of the light curve is seen when comparing the (O-C) for the data of JD1 and JD4 to that of the JD2 and JD3 data. Despite the incomplete sample of the JD1 and JD4 data, there is a strong resemblance between the amplitude of the residual of the JD1 and JD3 light curves and JD2 and JD4 light curves, suggesting that the brightness of the relevant active regions changes on timescales of day
 
The results for JD5 and JD6 are similar to those for JD2 in the overall shape and positioning of the features. We observe a small ($\Delta \phi \sim 0.05$) offset of the bright regions in the longitude of the two bright spots (from phase $\phi \sim 0.25$ in 1998 to phase $\phi \sim 0.3$ in 2000). Furthermore, we notice that the brighter of the two spots is the one centered at phase $\phi \sim 0.8$ in JD5 and JD6 whereas for JD2 the brighter spot is the one centered at phase $\phi \sim 0.25$.
Since the JD5 and JD6 data sets are not flux calibrated it is not possible to assess the correct flux level of the secondary star, which is essential for the fitting procedure. Therefore, caution must be exercised when comparing the results of the analysis of the \citetalias{2003MNRAS.345..506} and \citet{2002AJ....124.2853K}  data sets. JIn each individual data set it seems that active regions on this M star are stable and keep their location on time scales of, at least, a few days. "A comparison of the 1998 and 2000 data, suggests that these active regions may be relatively stable over much longer times, of the order of years. Two-spots configurations, 0.5 phases apart are commonly present on single dM active stars, however their location on the surface of the star evolves with time \citep{2006IAUJD...8E..56V}. What we witness in QS Vir is likely in accord with ``permanent active longitudes'' on RS CVn binaries and FK Comae stars, of active regions kept in place via tidal interactions of the two stellar components \citep{1998A&A...336L..25B}.

\subsection{Spectroscopy}
\label{subsec:anl_spec}

\begin{figure}
\epsscale{1.0}
\plotone{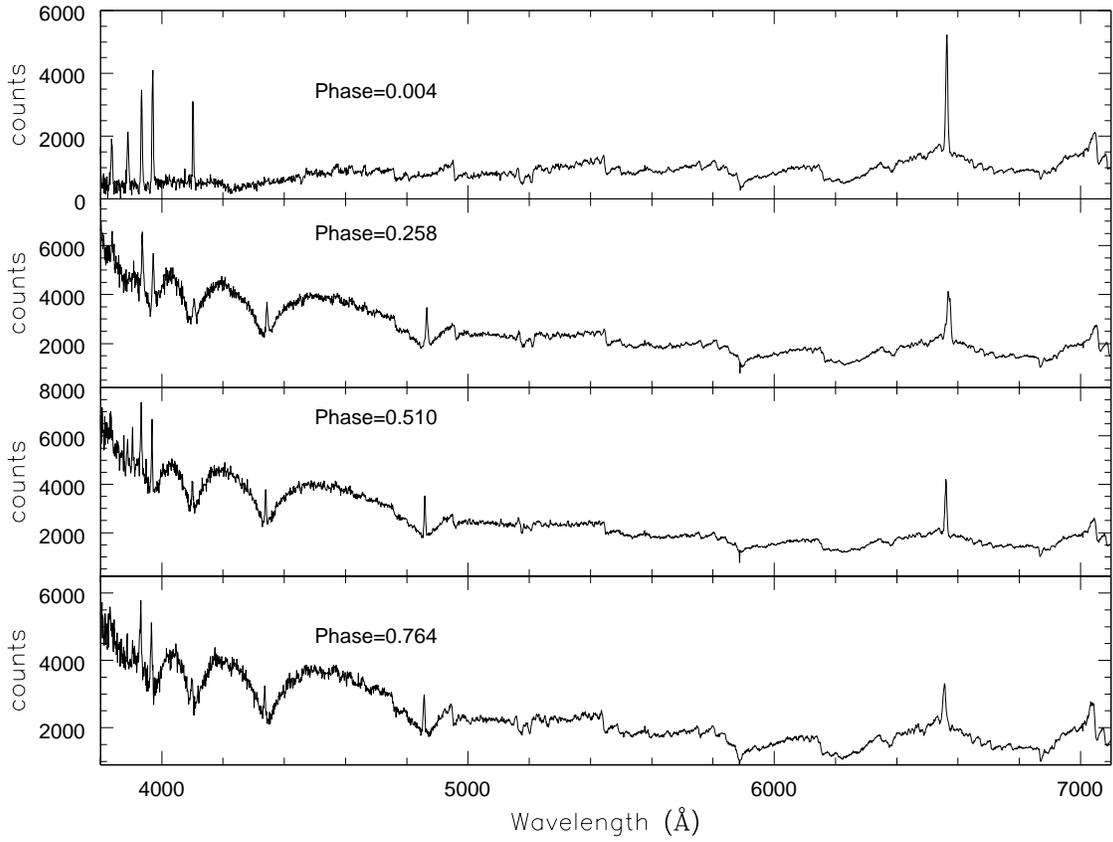}
\caption{Example spectra of QS Vir. At the top panel we show a spectrum taken during the eclipse, as can be noticed by the absence of the WD features .
}
\label{fig:main_spec}
\end{figure}

Fig. \ref{fig:main_spec} presents examples of our spectra, at four
different orbital phases (inferior/superior conjunction and
quadrature). The out-of-eclipse spectra are dominated by the
underlying WD in the blue and the M star in the red.
H$\alpha$ appears purely in emission along with the Ca~II H\&K
lines, whereas the other Balmer lines show narrow emission cores on
top of the broad absorption component from the WD. During eclipse, the
only emission lines present are the H$\alpha$ and the Ca$_{\rm II}$
H$\&$K lines, confirming that the latter and most of the former
originates on the M star (Fig. \ref{fig:main_spec}, top).

Fig. \ref{fig:trl_spec} presents trailed spectra for the Balmer and
Ca$_{\rm II}$ H\&K emission lines and for the TiO molecular band. Apart from the
H$\alpha$ emission line, all other Balmer emission lines  are
single-peaked and follow the motion of the M star
(Fig.~\ref{fig:trl_spec}). Their K velocities are presented in Table
\ref{tab:el_par}.  

The H$\alpha$ line appears to have a more complex structure, presenting three distinct components. These components are also present in our Echelle spectrum of Fig.~\ref{fig:Echelle}.  In Table \ref{tab:Echelle},  we summarize the properties of the emission line components obtained after a decomposition of the line profiles into gaussians. The different component are shown as dashed lines in Fig. \ref{fig:Echelle}  and are labeled accordingly. This spectrum was taken at a different epoch, indicating that the observed line structures are either persistent or at least very common for the H$\alpha$ line of QS Vir. The Echelle spectrum has a gap between  $4500 - 5500$\AA, therefore it is missing the H$\beta$ line. Notwithstanding, H$\gamma$ shows the same configuration as H$\alpha$ while only one extra feature is detectable on H$\delta$. It is quite likely that the lower resolution of the CTIO spectra prevents us from distinguishing multiple components in the other Balmer lines (as indicated by the structure seen at  $H\gamma$ in the Echelle spectrum). Possible mechanisms responsible for the asymmetric (and broadened) line profiles, is the Stark effect and flaring on the M star. Comparing our line profiles with the models of \cite{2005ApJ...630..573A} (also see Fig. 3 of \citealt{2006PASP...118..227}) we favor a flare on the M star as the mechanism that causes asymmetric line profiles and variations on the EWs in the Balmer lines of QS Vir.

The K velocity of the TiO band was measured by cross-correlating all spectra in the data set with a template spectrum to find the best-fit velocity displacement in each case. As previously noticed by \citetalias{2003MNRAS.345..506} and \citet{2005Ap&SS.296..481K}, the spectral type of the template star affects the resulting RVs amplitudes and leads to systematic differences which are comparable to the internal errors of the fit. In order to avoid this, we have used the spectrum of QS~Vir around mid-eclipse ($\phi$= 0.034) as our template. At these phases only the secondary star contributes to the observed spectrum as the WD is eclipsed. This method has the advantage of avoiding the uncertainties associated to the choices of spectral type for the secondary star, and determine the amount of line broadening one has to apply to a slowly-rotating template star. 

In addition to the selection of a correct template to correlate with the M star spectra, we have noticed that the choice of the spectral window has a strong impact in the resulting measured values. Specially, the selection of narrow spectral windows results in systematically lower values for the estimated K$_2$, while broader ranges results in larger values. We proceed implementing an iterative procedure that searches for the spectral range that minimizes the residue of the template spectra and spectra of the star corrected for the resulting radial velocity. This procedure results in K$_2 = 256 \pm 10$ km/s with a wavelength range of $\lambda = [6940^{+20}_{-5}:7250^{+10}_{-20}]$\AA. The uncertainty in the estimative of K$_2$ accounts for both the uncertainty in the fitting procedure and in the wavelength range selection. 

Our resulting K$_2 (= 256\pm 10$ km/s)  is consistent with the findings of \citetalias[(K$_2$ = 266 $\pm$ 6 km/s)]{2003MNRAS.345..506} and \citet[K$_2$ = 241$\pm$8 km/s]{2002AJ....124.2853K}. Adopting  our inferred parameters ($i$, $q$, K$_2$; see Table \ref{tab:params})  we find a binary solution with component masses of M$_1 = 0.6 \pm 0.3$ M$_\odot$ and M$_2 = 0.3 \pm 0.1$ M$_\odot$. The computed binary parameters are listed in Table \ref{tab:params}.

For the two nights of our spectroscopic monitoring, we measured the EW of the full H$\alpha$ line; the outcome is presented in Fig. \ref{fig:ews} and the relevant measurements are presented in Table \ref{tab:ew}. The four spectra of night 1 were taken around phase 0.1 and have EW of the full H$\alpha$ line close to 6\AA; this is similar to the \cite{2002AJ....124.2853K} quiescent H$\alpha$ EW of 5.6\AA. This measurement should represent a constant ``basal'' coronal emission from the M star, indicative of its activity level. On night 2, the EW of the line is highly variable, reaching $\sim$-18\AA~and -8\AA~at phase 0.0 in two successive cycles. It is very likely that, at the second night of our observations, we observed a flare on the M star, with maximum strength at inferior conjunction of the star. The EW modulation is likely a combination of a projection effect (due to the rotation of the system) and flare evolution, decreasing its magnitude by almost 50$\%$ in the next cycle. Furthermore, the EW of the star declines to -6\AA~between phases 0.5 and 0.75 but returns to a value of -8\AA~at inferior conjunction. This abrupt ``jump'' of the EW value after phase 0.75 suggests that we either witness the end of the previous event or the beginning of a new one, best visible at phase 0.0. Taking the EW variation as an indication of the extent of the active region that is responsible for the observed activity, we assess that it should be centered at the back side of the M star. The EW measurements of the full H$\alpha$ line are presented in Table \ref{tab:ew}.

\subsection{Doppler Imaging}
\label{subsec:anl_di}

An alternative representation of the emission line structure can be provided using Doppler tomography (DT) techniques. It uses the changes in line profile with binary phase   to compute a map of the line emitting sources in a two-dimensional   Doppler velocity space. The technique assumes that (i) all points   of the brightness distribution are equally visible at all phases;   (ii) there are no intrinsic changes in brightness with binary phase; (iii) all movement is parallel   to the orbital plane; (iv) all brightness sources are fixed in the   rotating binary frame; and (v) the intrinsic width of the line profile is negligible \citep{2001LNP...573....1M}. We apply the maximum entropy code provided by \cite{1998astro.ph..6141S} to produce Doppler maps for H$\alpha$, H$\beta$, H$\gamma$, H$\delta$ and also for the Ca$_{\rm II}$ H\&K doublet. For a more detailed description of the code, see \cite{1998astro.ph..6141S}. 

In order to make a direct comparison of our results with \citetalias{2003MNRAS.345..506} we have made the same assumptions with respect to the geometrical position of the WD and the M star.  Since DT does not take eclipses into account, spectra taken during eclipse were excluded from the image reconstruction procedure. The results for all emission lines are presented in Fig.~\ref{fig:dm1}. 

\begin{figure}

\epsscale{1.0}
\plotone{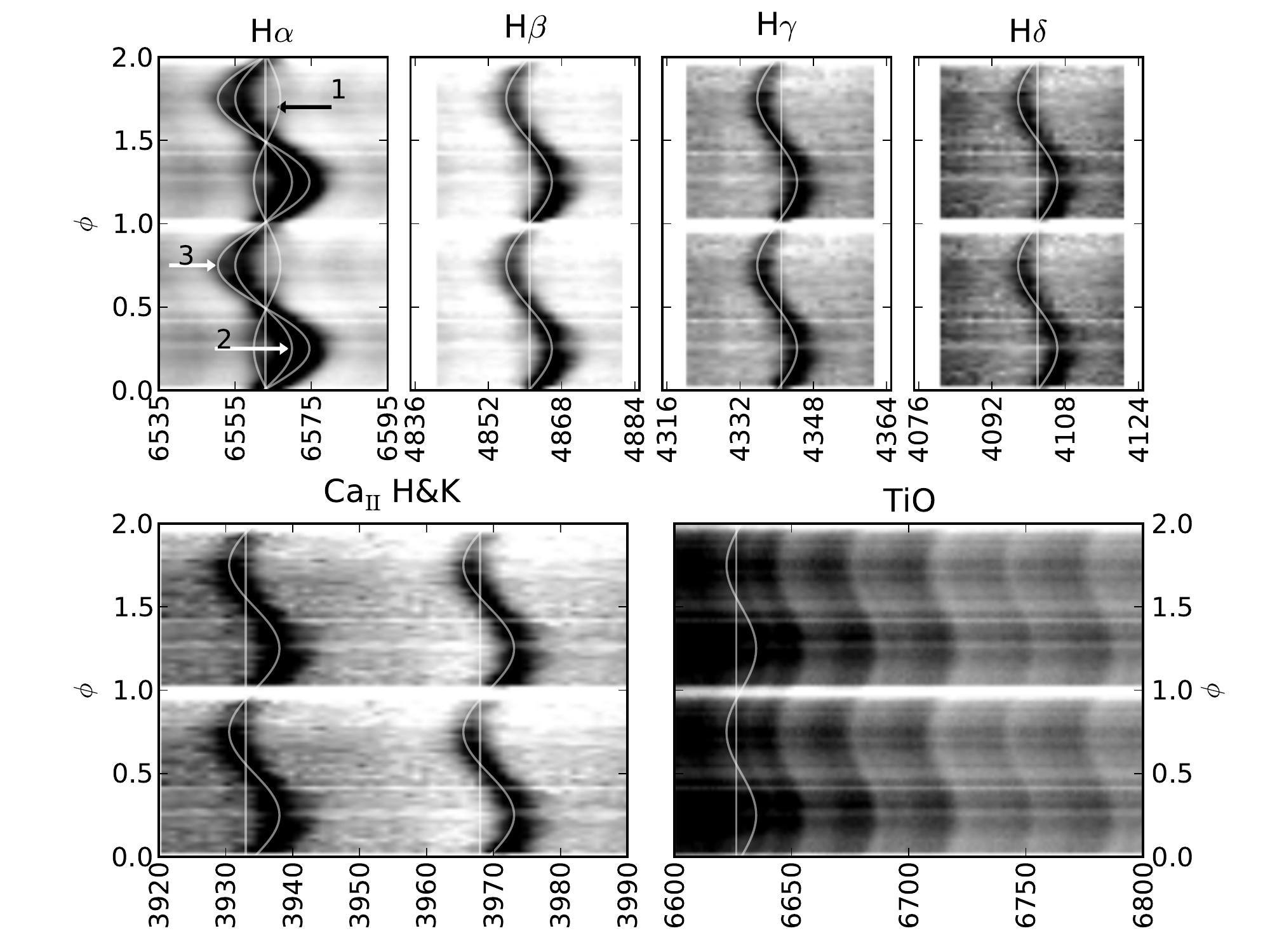}
\caption{Trailed spectra of the major Balmer lines (top four panels) and the Ca$_{\rm II}$ H\&K doublet and the TiO absorption band of QS Vir spectra. The position of the rest  wavelengths for each line/band and radial motion of the secondary are depicted as solid lines. 
}
\label{fig:trl_spec}
\end{figure}

\begin{figure}
\epsscale{1.0}
\plotone{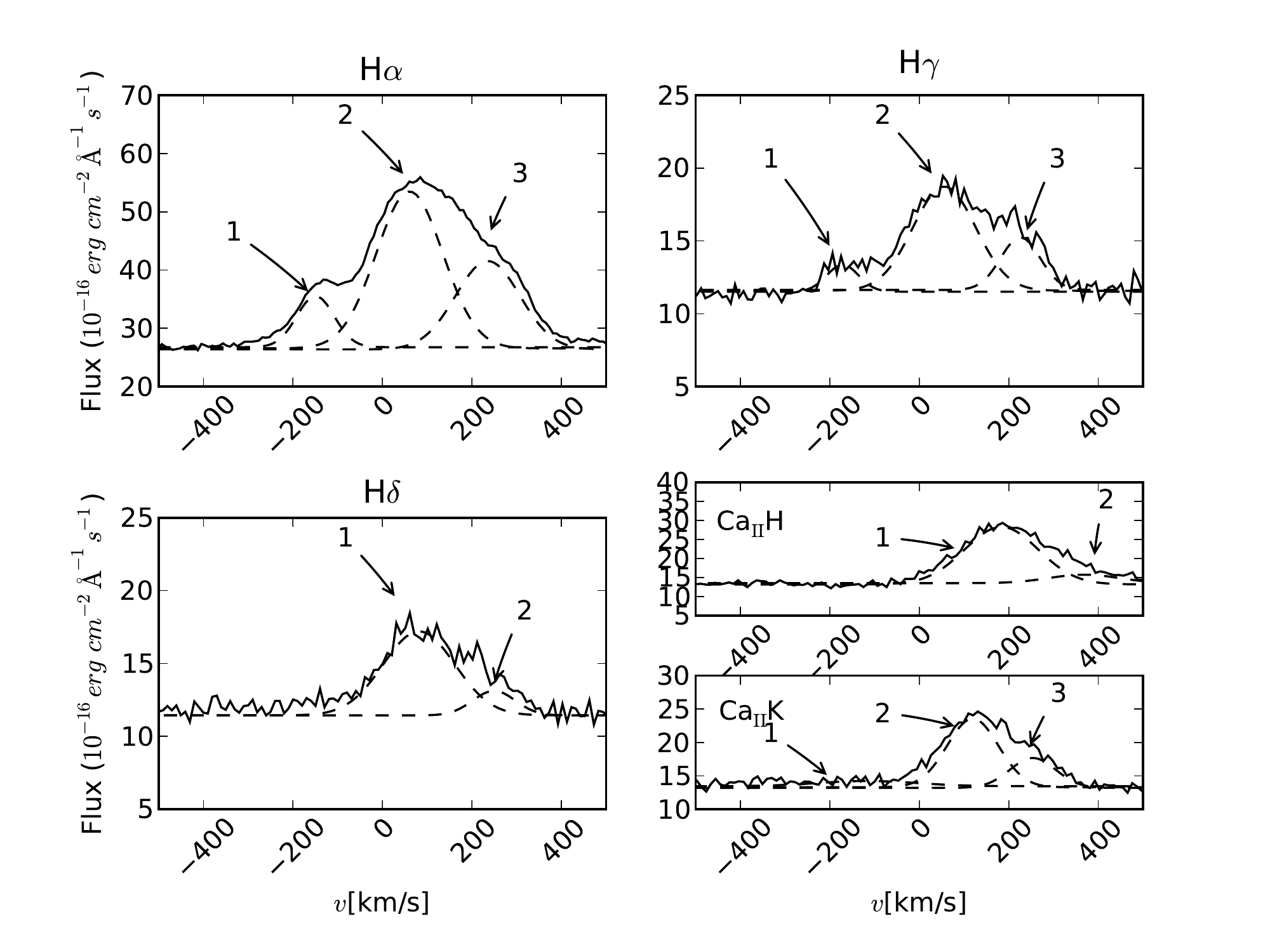}
\caption{Echelle spectra of Balmer and Ca$_{\rm II}$H\&K lines of QS Vir are shown as solid lines. In order to increase the S/N we have combined the data in bins of 0.1\AA. Dashed lines shows the resulting fit of Gaussians to the line profile with arrows indicating each component (according to the results in Table \ref{tab:Echelle}).}
\label{fig:Echelle}
\end{figure}

\begin{figure}
\epsscale{1.0}
\plotone{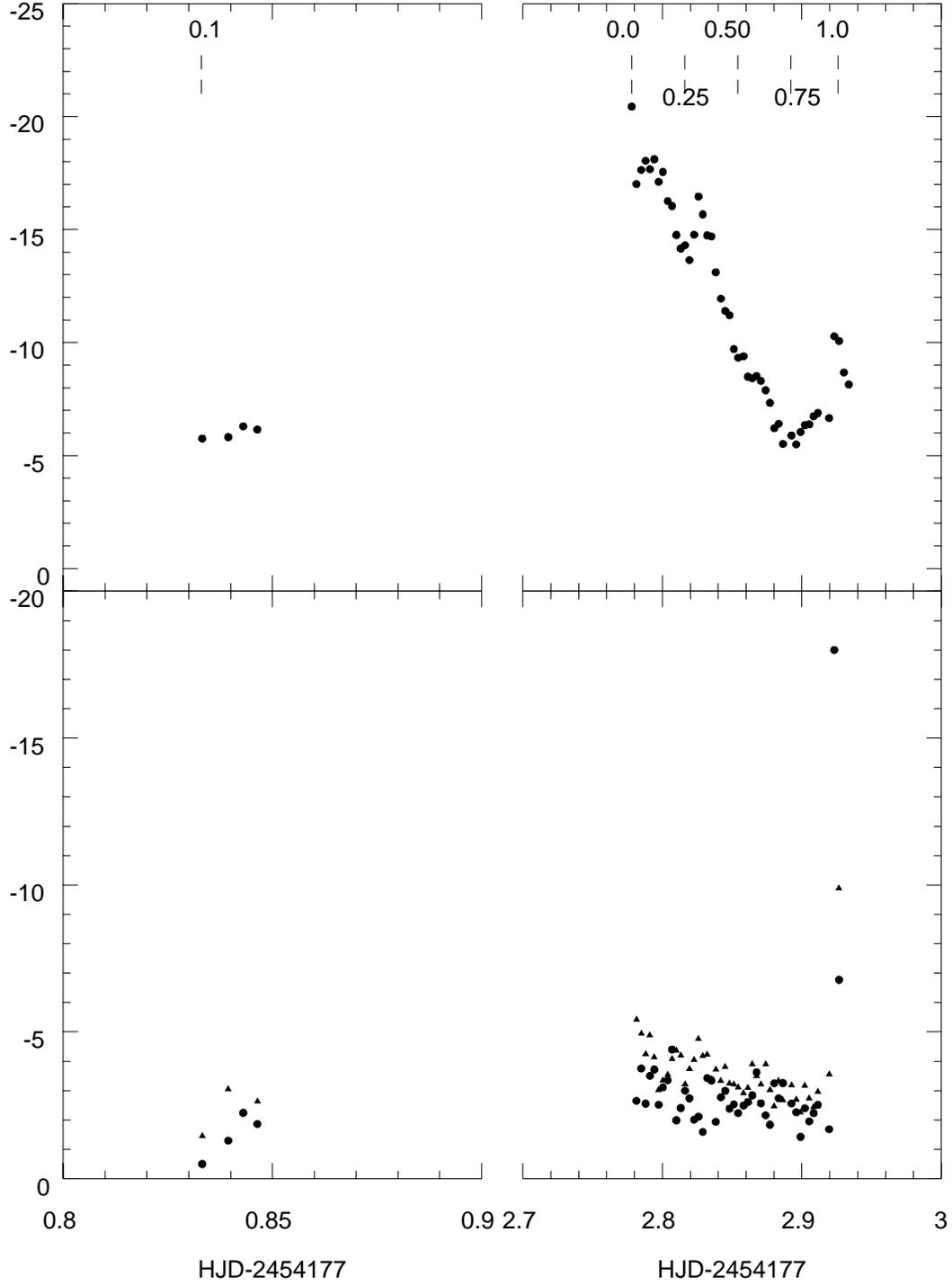}
\caption{EW of the H$\alpha$ line (top) and Ca$_{\rm II}$ H$\&$K (bottom) vs time. At the top figure, we mark the orbital phases of the system. See text for details.}
\label{fig:ews}
\end{figure}

\begin{figure}

\epsscale{1.0}
\plotone{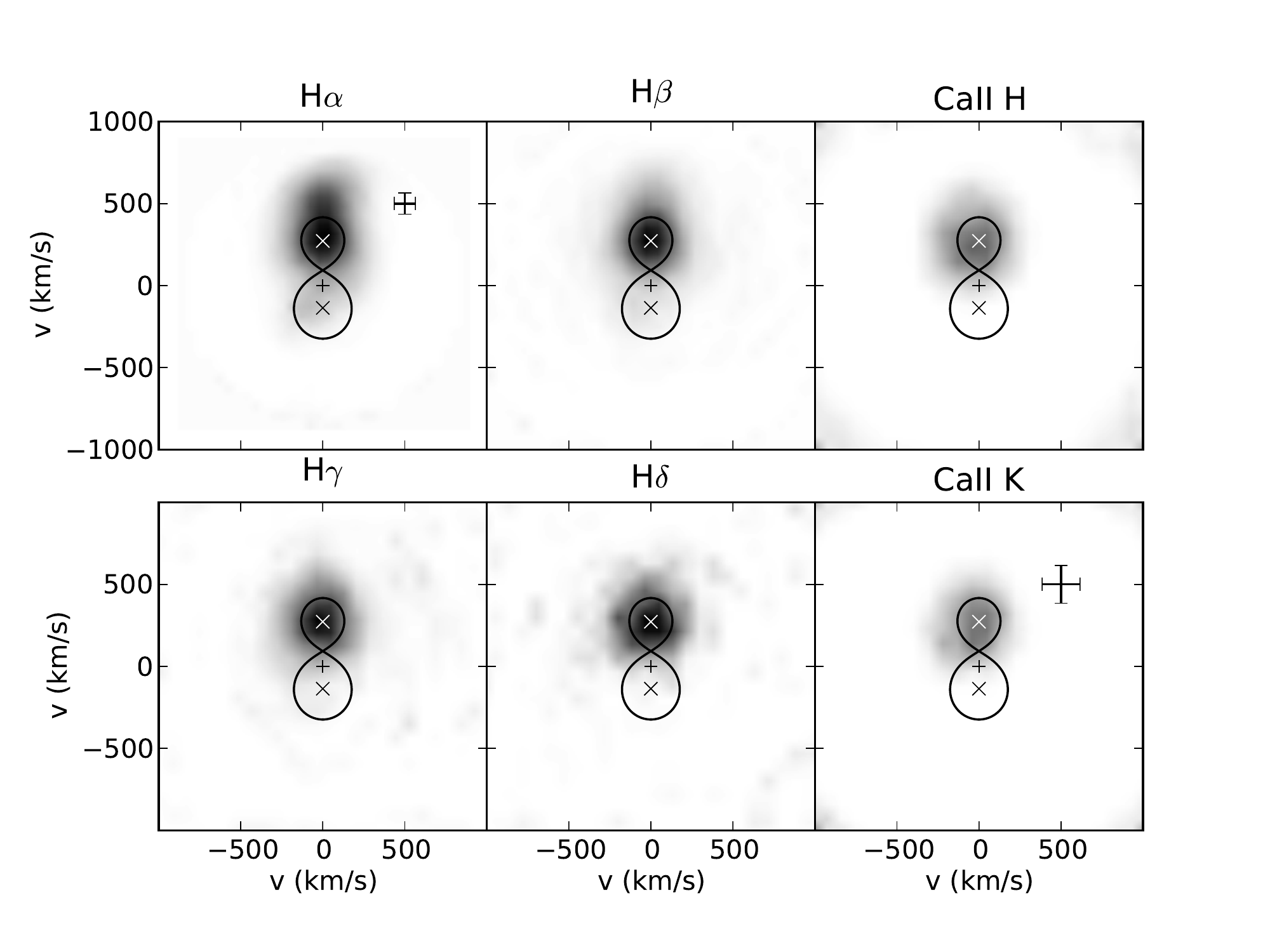}
\caption{Doppler tomograms of Balmer and Ca$_{\rm II}$ emission lines. Solid lines shows the Roche lobe for the parameters listed in Table \ref{tab:paramsdop} and crosses depicts the positions of the center of each component and of the center of mass of the system ($v = 0$). The black cross at the upper right panel of the H$\alpha$ and Ca$_{\rm II}$K maps represents the corresponding velocity-space resolution of the data.}
\label{fig:dm1}

\end{figure}

The most prominent feature in all Doppler maps is emission component centered on the secondary star. For the H$\alpha$ line, this feature corresponds to our component \#2 of the trailed spectrogram (Fig.~\ref{fig:trl_spec}). Our H$\alpha$ map also presents emission from inside the Roche lobe of the WD (this corresponds to our component \#1). The increased emission present at the back side of the secondary is the structure responsible for the component \#3 or "H$\alpha_{3}$" of the deconvolved spectra (see Fig.~\ref{fig:trl_spec} and Table~\ref{tab:el_par}). This indicates either that there is an additional source of emission at the back side of the M star, or that we observed the system when the M star was flaring, in agreement with our EW analysis conclusions of Fig. \ref{fig:ews}.

\begin{figure}

\epsscale{1.0}
\plotone{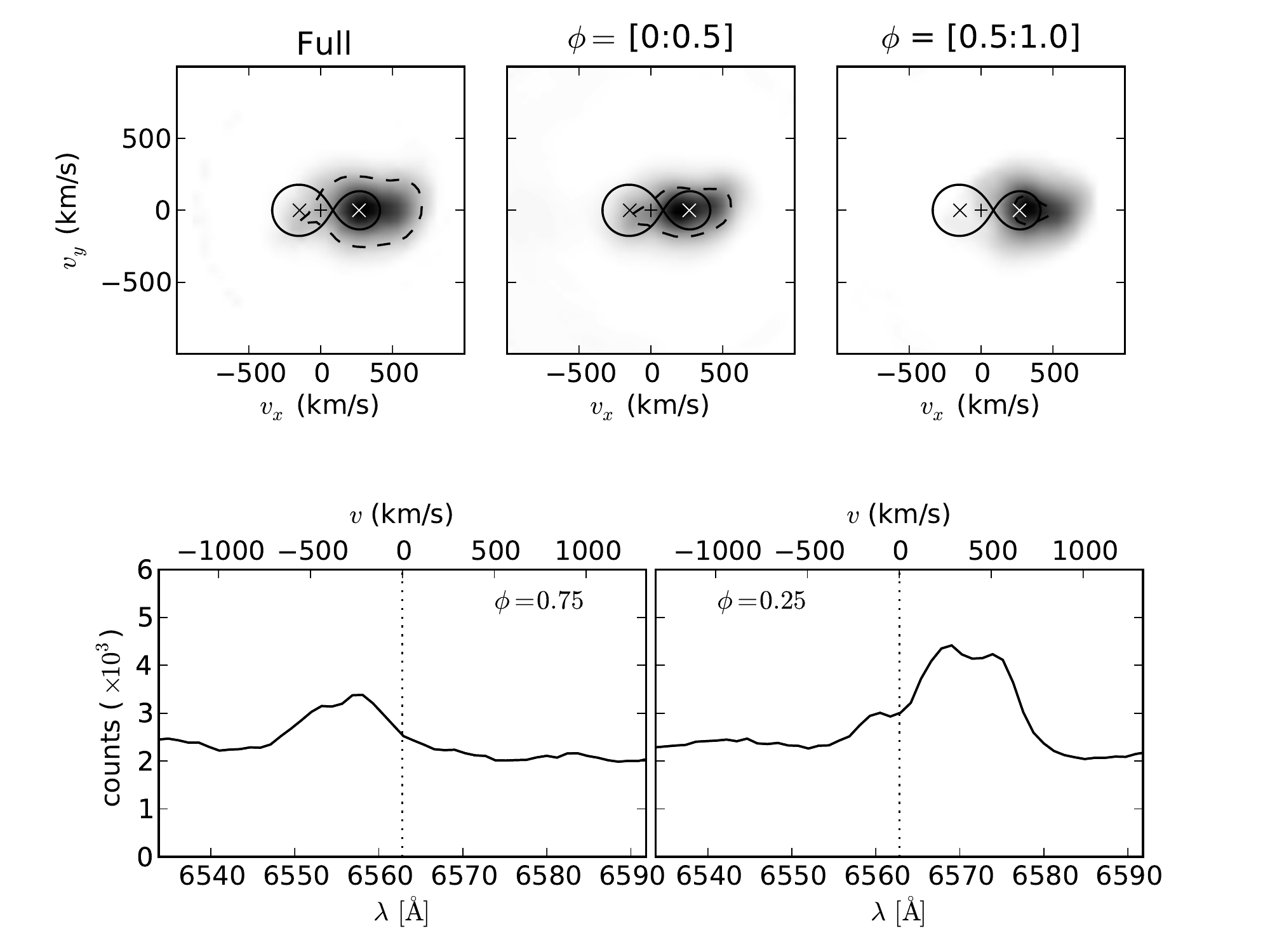}

\caption{{\bf Upper panel:} Doppler tomograms of H$\alpha$ comparing the map produced using all data for the second night of observations (Full) and for different half-cycles, of the same data set. A contour line depicts a region where the intensity is the same in all maps (set to the average of the middle panel map). {\bf Botton panel:} Example H$\alpha$ line profile separated by half a cycle. If all Doppler tomography assumptions were satisfied (see text for details), the profile should be a mirrored image. The dotted vertical line depicts the zero-velocity position, i.e. assuming $\lambda_0 ($H$\alpha) = 6562.8$\AA.}
\label{fig:rt_cicles}

\end{figure}

A useful test to assess the validity of the results from DT is to compare maps produced for different half cycles (from phases 0 to 0.5 and 0.5 to 1.0)\footnote{Here phases 0 and 1 are used only as reference. As we have stated earlier, spectra during eclipse were excluded from the imaging process.}; this helps us to estimate what regions of the system are satisfying DT assumptions (e.g. \citealt{1998MNRAS.294..689H, 2001LNP...573....1M}). The outcome of this exercise is presented in Fig. \ref{fig:rt_cicles}. The main difference observed for the first to second half cycle H$\alpha$ maps, is a reduction in the intensity of the corresponding features \#1 and \#3 of the trailed spectra (Figs. \ref{fig:trl_spec} and~\ref{fig:rt_cicles}). The bottom panel of Fig. \ref{fig:rt_cicles} shows the line profiles close to phases $\phi = 0.25$ and 0.75. If all assumptions of DT were satisfied, the two representations would be mirror images of each other (around zero velocity). However the low velocity component is almost absent in the second half cycle ($\phi = 0.75$) implying that it is either fading along the orbit or that it has an anisotropic emission. 

\section{Discussion}
\label{sec:disc}

It is challenging to pinpoint the source of origin and the mechanism for the various H$\alpha$ emission line components. Component \#1 is more intriguing, since it seems to change with time. Material inside the WD's Roche lobe was also present in the Doppler maps of \citetalias{2003MNRAS.345..506} and was perceived as a persistent structure. Although the Doppler map of \citetalias{2003MNRAS.345..506} and ours are similar\footnote{with the exception of the lack of feature \#3 in the \citetalias{2003MNRAS.345..506} data} the location of the emission component originating inside the Roche lobe of the WD appears to be different.  Since we have only one orbit of the binary (thus we can not assess orbital stability) this feature is either fading during our observations (therefore it is transient) or it has an anisotropic irradiation field and is seen preferentially at phase 0.25 compared to phase 0.75.  An inspection of the trailed spectra of \citetalias{2003MNRAS.345..506} (see their Fig. 15), reveals that the same structure is also much fainter close to $\phi \sim 0.75$ than $\phi\sim 0.25$. However their Doppler map was constructed using all the data from their spectroscopic observations (thus data from different epochs), which likely has smeared out any information about possible time evolution of the feature and/or its location in the WD Roche lobe at different times.

From our light curve models we derive a radius for the M star that is smaller than its Roche Lobe   radius at the 2-$\sigma$ confidence level (see Table \ref{tab:params}).  Therefore, accretion should not takes place through L1 as a result of standard Roche lobe overflow of the M star. At the same time, wind accretion is a possible scenario to explain material that lingers inside the Roche lobe of the WD, likely trapped by its gravitational potential. Active chromospheres of single M stars reach $\sim$1.1 stellar radii and are also easily blown up as a stellar wind \citep{2000MNRAS.316..699D}.  There is some evidence that mass loss via stellar wind of early-type M stars is of the order of $10^{-15} - 10^{-14}$ M$_\odot$yr$^{-1}$ (see \citealt{2006ApJ...652..636D} for a detailed discussion). The low temperature of the WD, compared to CVs of the same orbital period \citep{szk}, and lack of observational evidence for an accretion disc suggest that if there is accretion then it should be at a very low level. We conclude that a stellar wind should be the origin of the material inside the Roche lobe of the WD in QS Vir. Nevertheless, with such a low \.{M} it is hard to believe that any shock or collision will actually form from that material, which was the interpretation of \citetalias{2003MNRAS.345..506} for the features inside the WD Roche lobe, also observed in our H$\alpha$ trailed spectrum and Doppler maps. Material outside the Roche lobe  or inside the lobe of the WD cannot be in a stable orbital motion with the system unless it is kept there by some factor other than gravity. An interesting alternative could be a gas cloud locked with the M star by a magnetic prominence system, as has been reported by \cite{2000MNRAS.316..699D}.

Using a light curve inversion method we assess the details in the brightness distribution of the M star at six epochs of asymmetric light curves (using data from \citetalias{2003MNRAS.345..506} and \citealt{2002AJ....124.2853K}). In Figure \ref{fig:clt} we present our results for the epochs of observations that have good orbital coverage, namely JD2, JD3, JD5 and JD6. In JD2, there are two bright active regions centered at $\phi \sim 0.2$ and $\phi \sim 0.7$, which seem to be shifted slightly in phase ($\Delta \phi \sim 0.1$) in JD3. The JD3 map, in turn, is dominated by a strong flare; which will be commented later. From the partial light curves of JD1 and JD4, we have indications that these activity locations are also present during those epochs of observations. Interestingly, active regions at approximately the same region appears in JD5 and JD6. 

The stability of active regions with time prompts to the consistency of active longitudes in RS CVn and FK Comae systems.  \cite{2009MNRAS.395..282K} find that such stable active regions correspond to locations where magnetic flux tubes of opposite polarity emerge on the surface of the stars. Although similar high-resolution long-term data are not available for QS Vir, it will be interesting to investigate if such active regions evolve in latitude - as is suggested by the analysis of the JD5/JD6 data sets -  in synchronism with activity cycles of the M star, or if such cycles have no effect on the location of the activity regions in QS~Vir-type binaries. Possible implications on the level of magnetic braking in such systems (and consequently on the evolution of pre-CVs) remain to be determined.

 At JD3 the flare appears during eclipse and disappears half a cycle later as the binary rotates. The flare is likely located between $\phi \sim 0.2$ and $\phi \sim 0.4$ (Fig. \ref{fig:clt}), centered on $\phi \sim 0.3$. Furthermore, in our spectra, the EWs of the total H$\alpha$ emission line indicate that a flare was also present during our spectroscopic observations. The EWs analysis place the source of the variation at the back side of the M star, which is confirmed by the Doppler maps, reconstructed with the same data set. This does not come as a surprise, since this M star seems to be very active at all epochs of relevant observations (eg. \citetalias{2003MNRAS.345..506}). Regarding the decrement observed in the intensity of the Balmer lines  - i.e. the component been stronger in H$\alpha$ with decreasing intensity for the other Balmer components, we believe that the M star has experienced a solar-like flare \citep{2005A&A...444..593G}.  No similar structure at or close to that same region is found in the Doppler map of \citetalias{2003MNRAS.345..506}, supporting the suggestion that this is a transient feature.

A remarkable property of this system is its apparently high recurrence rate of flares which, in principle, could assist material finding its way to the WD. In the context of close detached binaries, there is evidence that tidal interactions can induce the emergence of magnetic flux tubes at preferred longitudes on the surface of the star \citep[for example]{2003A&A...405..303H}. With this scenario it is possible to explain the presence of material co-rotating with the binary within the Roche lobe of the WD in QS Vir as material trapped in magnetic loops from the M star. Changes of the position of the corresponding feature will occur accordingly with the regeneration/evolution of the gas cloud and the geometry of the magnetic field of the M star at that location. \cite{2000MNRAS.316..699D} measured a flux ratio of H$\alpha$/H$\beta \sim 10$ for the gas cloud of the post T Tauri star RX J1508.6-4423; our respective value is H$\alpha$/H$\beta \sim 14$ for the equivalent structure, in agreement with this interpretation. 

\section{Summary and conclusions}
\label{sec:end}

The main results of our time-resolved photometry and optical spectroscopy can be summarized as follows:

\begin{itemize}

\item  We provided light curve models of QS Vir, which (combined with the subsequent determination of K$_2$) constraining of the binary parameters. The size of the M star yields a distance to the system of $40\pm5$ pc. We also find that the M star radius is smaller than its Roche lobe radius at a 2-$\sigma$ confidense level which, therefore, suggests that accretion does not take place via Roche lobe overflow in QS Vir. The main binary parameters results are summarized in Table \ref{tab:params}.

\item Surface images of the M star were constructed and show four active regions with two different morphologies. There are two cool spotted regions 0.5 in phase apart and two hot active regions also 0.5 in phase apart. This configuration is persistent during a flare observed on the JD3 light curves and in the JD5/JD6 data. The different data sets JD1 to JD4 and JD5/JD6 are separated by $\sim 9$ yrs. The spot responsible for the flare on the JD3 data is centered at $\phi \simeq 0.3$ and occupies a sizable fraction of the star surface, with an azimuthal width of $\simeq 90 \degr$.

\item We have measured the RVs of the multi-component H$\alpha$ emission line and of other Balmer and Ca$_{\rm II}$ H\&K emission lines as well as of the TiO band. The results indicate that the chromospheric emission lines have larger K velocities than that expected for the centre of mass of the M star (as measured by the TiO band).  The results are listed in Table \ref{tab:el_par}.

\item Analysis of our time resolved spectroscopy and Doppler images of the H$\alpha$ emission line argue against the interpretation of the structures seen on the WD side of the binary as a result of low level accretion stream shock. It is likely that it is a result of wind accretion onto the WD. If the material inside the Roche lobe of
the WD is long-lived, we suggest that it can be held in place by a prominence-like 
magnetic loop in a manner similar to that observed on single active stars (see \citealt{2009MNRAS.395..282K} and references therein).

\end{itemize}

The detection of H$\alpha$ emission from the WD side of the binary was interpreted as a residual of a previous accretion state of QS Vir, suggestive that this is a hibernating CV \citepalias{2003MNRAS.345..506}. With the data in hand, we can not examine the long term spectroscopic behavior
of the system in order to compare the location and strength of the material inside the Roche lobe of the WD with time, and assess its stability and persistence. Our data also provide no indication that QS Vir bears characteristics of a hibernating CV. Since this is one of the few well-studied detached M dwarf-WD binaries, future observations complementing our work can assess the nature and stability of the active regions on the surface of the M star, and their role in angular momentum loss via magnetic braking, which drives the evolution of such systems.

\acknowledgments

We would like to thank D.W. Hoard for useful comments on an earlier draft of the manuscript. TR acknowledges financial support from CAPES. RB  acknowledges financial support from CNPq through grant 302.443/2008-8. 

This publication makes use of data products from the Two Micron All
Sky Survey, which is a joint project of the University of
Massachusetts and the Infrared Processing and Analysis
Center/California Institute of Technology, funded by the National
Aeronautics and Space Administration and the National Science
Foundation.

Facilities: \facility{Blanco 4-m, VLT 8-m, OPD 1.6-m}

\clearpage

\clearpage

\clearpage

\clearpage

\clearpage

\clearpage

\clearpage

\clearpage

\clearpage


\begin{deluxetable}{cccccccc}
\tabletypesize{\scriptsize}
\tablecaption{\label{tab:obs_log} Log of observations}
\tablewidth{0pt}
\tablehead{
\colhead{Obs. type} & \colhead{Date} & \colhead{Filter} & \colhead{No exposures} & \colhead{Exp. time} & \colhead{ndith} & \colhead{Start} & \colhead{End}}
\startdata
Photometry   & 18 03 2008 & $J$ & $53$& $45s$ & $4$ & 2 454 544.591 & 2 454 544.659 \\
Photometry   & 18 03 2008 & $H$ & $49$& $30s$ & $4$ & 2 454 544.595 & 2 454 544.660 \\
Spectroscopy & 20 03 2007 &     & $46$& $180s$& --  &  2 454 177.833 & 2 454 177.846 \\
Echelle & 16 07 2007 &     & $1$& $1800s$& --  &  2 454 329.538 & -- \\
\enddata
\end{deluxetable}

\begin{deluxetable}{cccccc}

\tablecaption{\label{tab:params} Model Parameters from the light curve fits and resulting binary parameters. The radius of the the secondary star ($r_2$) and its volumetric radius ($r_{LR}$, \citealt{1983ApJ...268..368E}) are given in units of orbital separation.}
\tablewidth{0pt}
\tablehead{\colhead{Parameter} & \colhead{V} & \colhead{R} & \colhead{I} & \colhead{J} & \colhead{H}}
\startdata
$F_2(mJy)$ & $3.05\pm0.02$ &$7.66\pm 0.04$ & $28.2\pm 0.1$&$68.3\pm 0.4$ &$72.8\pm 0.4$ \\ 
$F_1(mJy)$ & $3.42\pm0.02$ &$2.60\pm 0.05$ & $2.2\pm 0.2$ &$0.9\pm 0.1$  &$1.0\pm 0.2$ \\
$\chi^2$ & $2.74$ & $2.62$& $1.20$ & $1.05$  & $0.94$ \\
\hline 
\multicolumn{6}{c}{Simultaneously fitted parameters:} \\
\multicolumn{2}{c}{$i = (74.9\pm0.6)^\circ$} & \multicolumn{2}{c}{$q =(0.50\pm0.05)$}  & \multicolumn{2}{c}{$r_{LR} = (0.44 \pm 0.03)$}  \\
\multicolumn{3}{c}{$r_2 = (0.36^{+0.02}_{-0.06})$} & \multicolumn{3}{c}{$r_2/r_{\rm RL} = (0.82^{+0.07}_{-0.2})$} \\
\hline
\multicolumn{6}{c}{Binary parameters:} \\
\multicolumn{2}{c}{M$_1 = 0.6\pm0.3$ M$_\odot$} & \multicolumn{2}{c}{M$_2 = 0.3\pm0.1$ M$_\odot$} & \multicolumn{2}{c}{R$_2 = 0.43^{+0.04}_{-0.07}$ R$_\odot$} \\
 \multicolumn{3}{c}{d = $40\pm5$ pc}  &  \multicolumn{3}{c}{ $a = 1.2\pm0.1$ R$_\odot$} \\
\enddata

\end{deluxetable}

\begin{deluxetable}{cccc}
\tablecaption{\label{tab:el_par} Parameters of emission lines}
\tablewidth{0pt}
\tablehead{
\colhead{line} & \colhead{$\gamma$ (km/sec)} & \colhead{K (km/sec)} & \colhead{$\phi_{0}$}}
\startdata
H$\alpha_{1}$    & ~18.6$\pm$0.7   & 158$\pm$18 &0.08$\pm$0.02 \\
H$\alpha_{2}$     & -34.7$\pm$0.9  & 307$\pm$8 & 0.462$\pm$0.004\\
H$\alpha_{3}$ & -23$\pm$2  & 549$\pm$4 &  0.457$\pm$0.004\\
H$\beta$   & -9.2$\pm$0.6  & 307$\pm$5 & 0.462$\pm$0.003\\
H$\gamma$  & -61.5$\pm$0.6 & 297$\pm$5 & 0.450$\pm$0.003 \\
H$\delta$  & -20.9$\pm$0.9 & 285$\pm$10 & 0.467$\pm$0.006 \\
Ca$_{\rm II}$H      & ~53.4$\pm$0.2  & 278$\pm$4 &0.452$\pm$0.003 \\
Ca$_{\rm II}$K      & 14.2$\pm$0.2  & 281$\pm$5 & 0.451$\pm$0.003  \\
TiO                           &  -- & $256 \pm 10$ & -- \\
\enddata
\end{deluxetable}

\begin{deluxetable}{ccccc}
\tablecaption{\label{tab:Echelle} Parameters of our Echelle spectra emission lines.}
\tablewidth{0pt}
\tablehead{\colhead{Component} & \colhead{$v$ [km/s]} & \colhead{EW} & \colhead{Flux}}
\startdata
H$\alpha$ - 1 & -158 & -5.19 & 32.5 \\
H$\alpha$ - 2 & 51     & -1.25 & 133 \\
H$\alpha$ - 3 & 226  & -3.00 & 76.7 \\
H$\gamma$ - 1 & -171 & -0.24 & 2.86 \\
H$\gamma$ - 2 & 56   & -1.89& 21.9\\
H$\gamma$ - 3 & 226  & -0.69 & 6.97 \\
H$\delta$ - 1 & 104 & -1.66 & 19.0 \\
H$\delta$ - 2 & 249 & -0.28 & 3.30 \\
Ca$_{\rm II}$ H -  1& 183  & -3.7  & 49.4 \\
Ca$_{\rm II}$ H - 2 & 368  & -2.93 & 6.90 \\
Ca$_{\rm II}$ K -  1& -116  & -0.25  & 3.45 \\
Ca$_{\rm II}$ K -  2& 119  & -1.75  & 23.2 \\
Ca$_{\rm II}$ K - 3 & 253  & -0.69 & 9.06\\

\enddata
\end{deluxetable}

\begin{deluxetable}{ccc}
\tablecaption{\label{tab:ew} EWs for the H$\alpha$ line}
\tablehead{\colhead{HJD} & \colhead{phase} & \colhead{EW [\AA]} }
\startdata 
2454177.8332  &  0.1039  &  -5.75  \\
2454177.8395  &  0.1452  &  -5.82  \\
2454177.8430  &  0.1689  &  -6.30  \\
2454177.8464  &  0.1913  &  -6.15  \\
2454179.7780  &  0.0042  &  -20.44  \\
2454179.7814  &  0.0267  &  -17.01  \\
2454179.7849  &  0.0494  &  -17.64  \\
2454179.7880  &  0.0700  &  -18.04  \\
2454179.7911  &  0.0906  &  -17.67  \\
2454179.7942  &  0.1112  &  -18.11  \\
2454179.7973  &  0.1318  &  -17.11  \\
2454179.8004  &  0.1524  &  -17.55  \\
2454179.8038  &  0.1753  &  -16.26  \\
2454179.8069  &  0.1959  &  -16.04  \\
2454179.8100  &  0.2165  &  -14.76  \\
2454179.8132  &  0.2371  &  -14.16  \\
2454179.8163  &  0.2577  &  -14.31  \\
2454179.8194  &  0.2783  &  -13.65  \\
2454179.8228  &  0.3008  &  -14.77  \\
2454179.8259  &  0.3214  &  -16.46  \\
2454179.8290  &  0.3420  &  -15.67  \\
2454179.8321  &  0.3626  &  -14.74  \\
2454179.8352  &  0.3832  &  -14.69  \\
2454179.8383  &  0.4038  &  -13.11  \\
2454179.8419  &  0.4279  &  -11.94  \\
2454179.8450  &  0.4485  &  -11.40  \\
2454179.8481  &  0.4691  &  -11.21  \\
2454179.8512  &  0.4897  &  -9.71  \\
2454179.8543  &  0.5103  &  -9.33  \\
2454179.8582  &  0.5361  &  -9.39  \\
2454179.8613  &  0.5567  &  -8.49  \\
2454179.8644  &  0.5773  &  -8.42  \\
2454179.8675  &  0.5979  &  -8.52  \\
2454179.8707  &  0.6185  &  -8.31  \\
2454179.8740  &  0.6409  &  -7.89  \\
2454179.8771  &  0.6615  &  -7.33  \\
2454179.8802  &  0.6821  &  -6.21  \\
2454179.8834  &  0.7027  &  -6.41  \\
2454179.8865  &  0.7233  &  -5.51  \\
2454179.8926  &  0.7640  &  -5.89  \\
2454179.8960  &  0.7865  &  -5.50  \\
2454179.8991  &  0.8071  &  -6.04  \\
2454179.9022  &  0.8277  &  -6.34  \\
2454179.9053  &  0.8483  &  -6.38  \\
2454179.9084  &  0.8689  &  -6.74  \\
2454179.9115  &  0.8895  &  -6.88  \\
2454179.9196  &  0.9432  &  -6.66  \\
2454179.9232  &  0.9672  &  -10.28  \\
2454179.9267  &  0.9901  &  -10.07  \\
2454179.9301  &  0.0132  &  -8.68  \\
2454179.9336  &  0.0363  &  -8.14  \\
\enddata
\end{deluxetable}

\begin{deluxetable}{cc}

\tablecaption{\label{tab:paramsdop} Model Parameters used to calculate the position of the centre of mass of both components and to draw the Roche lobe on the Doppler maps.}
\tablewidth{0pt}
\tablehead{\colhead{parameter} & \colhead{value} }
\startdata
i &  75.5\degr  \\
M$_{WD}$ & 0.78 M$_{\odot}$     \\
M$_{2}$ &  0.43 M$_{\odot}$    \\
P$_{orb}$ & 0.15075 d  \\
\enddata

\end{deluxetable}

\end{document}